\documentclass[aps,pra,twocolumn,showpacs,amsmath,amssymb,10pt]{revtex4-1}
\usepackage{graphicx}% Include figure files
\usepackage{bm}% bold math
\usepackage{slashed}
\usepackage{color}

\begin{document}
\title{Generalized Klein-Nishina formula}
\author{K. Krajewska$^{1,2}$}
\email[E-mail address:\;]{Katarzyna.Krajewska@fuw.edu.pl}
\author{F. Cajiao V\'elez$^1$}
\author{J. Z. Kami\'nski$^1$}
\affiliation{$^1$Institute of Theoretical Physics, Faculty of Physics, University of Warsaw, Pasteura 5,
02-093 Warszawa, Poland\\
$^2$Department of Physics and Astronomy, University of Nebraska, Lincoln, Nebraska 68588-0299, USA
}
\date{\today}
\begin{abstract}
The generalized Klein-Nishina formula for Compton scattering of charged particles by a finite train of pulses is derived in the framework of quantum electrodynamics. 
The formula also applies to classical Thomson scattering provided that frequencies of generated radiation are smaller that the cut-off 
frequency. The validity of the formula for incident pulses of different durations is illustrated by numerical examples. 
The positions of the well-resolved Compton peaks, with the clear labeling by integer orders, opens up the possibility of the 
precise diagnostics of properties of relativistically intense, short laser pulses. This includes their peak intensity, the carrier-envelope phase,
and their polarization properties.
\end{abstract}
\maketitle

\section{Introduction}
\label{sec:intro}

In order to understand the physics behind the interaction of strong laser pulses with matter, it is necessary to have well-characterized interacting pulses.
With current technology, this is accomplished for laser fields in the optical regime provided that their intensity is no larger than 10$^{15}$W/cm$^2$.
As pointed out by many authors (see, for instance, Refs.~\cite{Yanovsky,Link,Bahk,Hetzheim,Gao,Macken}), the same task becomes particularly challenging at
higher intensities, where any direct measurement is prone to damaging the equipment. Therefore, different proposals were put forward to determine
properties of ultra-strong and short laser pulses (e.g., the laser peak intensity~\cite{Yanovsky,Link,Bahk,Hetzheim,Gao} or the carrier-envelope phase~\cite{Macken}). 
As we will argue in this paper, a sensitivity of Compton scattering to the driving train of pulses can be traced back to the properties of pulses comprising the train.
This can be based on the {\it generalized Klein-Nishina (GKN) formula}, which we derive in this paper.

\subsection{Klein-Nishina formula}

The original Klein-Nishina formula~\cite{KleinNishina} for the Compton scattering with the electron initial four-momentum $p_{\mathrm{i}}$ [$p_{\mathrm{i}}\cdot p_{\mathrm{i}}=(m_{\mathrm{e}}c)^2$, 
where `\textit{dot}' means the relativistic scalar product, $a\cdot b=a^0b^0-a^1b^1-a^2b^2-a^3b^3=a^0b^0-\bm{a}\cdot\bm{b}$], has the form
\begin{equation}
\omega_{\bm{K}}=\frac{\omega}{\frac{p_{\mathrm{i}}\cdot n_{\bm{K}}}{p_{\mathrm{i}}\cdot n}+\frac{\omega}{\omega_{\mathrm{cut}}}},
\label{kn1}
\end{equation}
where $k=(\omega/c)n$ and $K=(\omega_{\bm{K}}/c)n_{\bm{K}}$ are the four-momenta of the initial and final photons, respectively, 
$n=(1,\bm{n})$, $n_{\bm{K}}=(1,\bm{n}_{\bm{K}})$, and $k\cdot k=K\cdot K=0$. In this equation,
\begin{equation}
\omega_{\mathrm{cut}}=\frac{cp_{\mathrm{i}}\cdot n}{\hbar n\cdot n_{\bm{K}}}
\label{kn8}
\end{equation}
is the maximum frequency of photons generated in the Compton process.

Consider the Compton process which takes place in an intense monochromatic plane wave of the frequency $\omega$, propagating in the direction $\bm{n}$. 
The vector potential of, in general, the elliptically polarized plane wave equals
\begin{equation}
A(k\cdot x)=A_0(\varepsilon_1\cos(k\cdot x)\cos\delta+\varepsilon_2\sin(k\cdot x)\sin\delta),
\label{kn3}
\end{equation}
where $\varepsilon_j$ ($j=1,2$) are two real polarization four-vectors normalized such that $\varepsilon_j\cdot\varepsilon_{j'}=-\delta_{jj'}$ and $k\cdot \varepsilon_j=0$.
Without the loss of generality, we assume that the time-components of these vectors are zero, i.e., $\varepsilon_j=(0,\bm{\varepsilon}_j)$.
If the electron absorbs $N=1,2,\ldots$ photons from the plane wave it may emit, in the direction $\bm{n}_{\bm{K}}$, the Compton photon of frequency 
$\omega_{\bm{K},N}$~\cite{BrownKibble,RitusNikishov,Ritus,EKK2009,PMHK2012,Ras}, 
\begin{equation}
\omega_{\bm{K},N}=\frac{N\omega}{\frac{p_{\mathrm{i}}\cdot n_{\bm{K}}}{p_{\mathrm{i}}\cdot n}+\frac{U}{c}\frac{n\cdot n_{\bm{K}}}{p_{\mathrm{i}}\cdot n}+\frac{N\omega}{\omega_{\mathrm{cut}}}},
\label{kn2}
\end{equation}
where
\begin{equation}
U=\frac{1}{4}\mu^2\frac{(m_{\mathrm{e}}c^2)^2}{cp_{\mathrm{i}}\cdot n}
\label{kn4}
\end{equation}
has the meaning of the ponderomotive energy of electrons in the laser field. Here, the relativistically invariant parameter,
\begin{equation}
\mu=\frac{|e|A_0}{m_{\mathrm{e}}c},
\label{kn5}
\end{equation}
determines the intensity of the electromagnetic plane wave. Both $U$ and $\mu$ are the classical quantities, whereas the term proportional 
to $\omega/\omega_{\mathrm{cut}}$ in \eqref{kn2} accounts for the quantum recoil of electrons during the Compton process. 
Such a recoil of electrons does not take place in the corresponding classical process, that is called the Thomson scattering. 
This allows to introduce into the Klein-Nishina formula the classical Thomson frequency, 
\begin{equation}
\omega^{\mathrm{Th}}_{\bm{K},N}=\frac{N\omega}{\frac{p_{\mathrm{i}}\cdot n_{\bm{K}}}{p_{\mathrm{i}}\cdot n}+\frac{U}{c}\frac{n\cdot n_{\bm{K}}}{p_{\mathrm{i}}\cdot n}},
\label{kn6}
\end{equation}
such that
\begin{equation}
\omega_{\bm{K},N}=\frac{\omega^{\mathrm{Th}}_{\bm{K},N}}{1+\frac{\omega^{\mathrm{Th}}_{\bm{K},N}}{\omega_{\mathrm{cut}}}}.
\label{kn7}
\end{equation}
Note that, for a given geometry of the process and for a given initial electron momentum, the Thomson frequencies are equally separated from each other. 
The same is not true for the Compton frequencies. This scaling law, which relates the quantum Compton frequency $\omega_{\bm{K},N}$ to its classical analog, 
the Thomson frequency $\omega^{\mathrm{Th}}_{\bm{K},N}$, has been discussed recently for long laser pulses \cite{SKscale1,SKscale2,SKscale3}. 
Its extension to short laser pulses together with the investigation of polarization and spin effects, and the synthesis of ultra-short pulses have been presented in~\cite{KKscaling}.

The Klein-Nishina formula \eqref{kn2} can be also expressed in the form
\begin{equation}
N=\frac{\omega_{\bm{K},N}\omega_{\mathrm{cut}}}{\omega (\omega_{\mathrm{cut}}-\omega_{\bm{K},N})}
\Bigl(\frac{p_{\mathrm{i}}\cdot n_{\bm{K}}}{p_{\mathrm{i}}\cdot n}+\frac{U}{c}\frac{n\cdot n_{\bm{K}}}{p_{\mathrm{i}}\cdot n}\Bigr),
\label{kn9}
\end{equation}
which, for the given geometry of the scattering process and for the given frequency of the emitted Compton photon, allows to determine the nonlinearity of the process, $N$. 
Note that the quantum nature of this formula is hidden in $\omega_{\mathrm{cut}}$. Therefore, we recover the classical result in the limit when $\omega_{\bm{K},N} \ll \omega_{\mathrm{cut}}$. 
Similar formulas can be derived for multichromatic plane-wave-fronted fields with commensurate frequencies.

\subsection{Spectrally resolved Compton peaks induced by short laser pulses}

If the Compton process occurs in a short and intense laser pulse, the situation is different. It follows from the time-frequency uncertainty principle that, if the driving pulse lasts for time 
$T_{\mathrm{p}}$, the frequency scale over which the system undergoes a significant change cannot be smaller than roughly $2\pi/T_{\mathrm{p}}$. 
For this reason, the individual peaks in the Compton frequency spectrum are hardly visible if the process occurs in few-cycle pulses (see, e.g., Refs.~\cite{SKscale1,SKscale2,KKcompton,BocaFlorescu,Mackenroth}). 
Now, the question arises: Is it possible to design a short laser pulse such that the individual peaks in the laser-induced Compton spectrum are clearly distinguished 
from each other, with unambiguously prescribed to them integer orders $N$? Although the answer to this question is in general negative, 
for suitably designed pulses, one can achieve the limit imposed by the aforementioned uncertainty relation. The idea of how to avoid the spectral broadening 
follows from the Fraunhoffer diffraction theory as applied to the diffraction gratings, or from the frequency comb generation, 
and is based on the application of modulated laser pulses \cite{KKcomb1,KKcomb2}. We will demonstrate in this paper that, by using this technique, it is possible 
to achieve the clear spectral resolution of individual peaks in the Compton spectrum, even if driven by few-cycle laser pulses.

The Compton scattering by short laser pulses is very sensitive to the precise form of the driving pulse.
It was pointed out in~\cite{Macken} that this sensitivity can be used to characterize the driving laser fields.
In Ref.~\cite{Macken}, the angular distributions of Compton radiation were traced back to the carrier-envelope phase of the driving pulse. 
As we will point out, the spectrally resolved peaks in the Compton distribution allow for determining properties 
of the driving train of pulses. This is done by mapping the positions of the Compton peaks to the GKN formula, 
that we derive in this paper for an arbitrary, finite train of pulses.

The paper is organized as follows. In the next section, the theory of Compton scattering in finite plane-wave-fronted pulses is presented, 
which is then followed by the discussion of the diffraction formula (Sec.~\ref{sec:diffraction}). In Sec.~\ref{sec:gklein}, we present 
the GKN formula. It describes the major peaks in the Compton spectrum 
of radiation induced by a finite train of pulses. Since the formula depends on properties 
of the individual pulse from the train, one may exploit the properties of the Compton spectra 
in the diagnostics of relativistically intense, short laser pulses. This is illustrated in Sec.~\ref{sec:short} for one- 
and three-cycle pulses. Sec.~\ref{sec:conclusions} contains concluding remarks.

In analytic formulas we keep $\hbar=1$ and, hence, the fine-structure constant becomes $\alpha=e^2/(4\pi\varepsilon_0c)$. 
Unless stated otherwise, in numerical analysis we use relativistic units (rel. units) such that $\hbar=m_{\rm e}=c=1$, where $m_{\rm e}$ is the electron mass.

\section{Compton scattering}
\label{sec:compton}

The probability amplitude for the Compton process, $e^-_{\bm{p}_{\mathrm{i}}\lambda_{\mathrm{i}}}\rightarrow e^-_{\bm{p}_{\mathrm{f}}\lambda_{\mathrm{f}}}+\gamma_{\bm{K}\sigma}$, with the initial and final electron momenta and spin polarizations $\bm{p}_{\mathrm{i}}\lambda_{\mathrm{i}}$ and $\bm{p}_{\mathrm{f}}\lambda_{\mathrm{f}}$, respectively, equals \cite{KKcompton}
\begin{equation}
{\cal A}(e^-_{\bm{p}_{\mathrm{i}}\lambda_{\mathrm{i}}}\rightarrow e^-_{\bm{p}_{\mathrm{f}}\lambda_{\mathrm{f}}}
+\gamma_{\bm{K}\sigma})=-\mathrm{i}e\int \mathrm{d}^{4}{x}\, j^{(++)}_{\bm{p}_{\mathrm{f}}\lambda_{\mathrm{f}},
\bm{p}_{\mathrm{i}}\lambda_{\mathrm{i}}}(x)\cdot A^{(-)}_{\bm{K}\sigma}(x). \label{ComptonAmplitude}
\end{equation}
Here, $\bm{K}\sigma$ denotes the Compton photon momentum and polarization, and
\begin{equation}
A^{(-)}_{\bm{K}\sigma}(x)=\sqrt{\frac{1}{2\varepsilon_0\omega_{\bm{K}}V}} 
\,\varepsilon^*_{\bm{K}\sigma}\mathrm{e}^{\mathrm{i}K\cdot x},
\label{per}
\end{equation}
where $V$ is the quantization volume, $\omega_{\bm{K}}=cK^0=c|\bm{K}|$ ($K\cdot K=0$), and $\varepsilon_{\bm{K}\sigma}=(0,\bm{\varepsilon}_{\bm{K}\sigma})$ are the polarization four-vectors satisfying the conditions
$K\cdot\varepsilon_{\bm{K}\sigma}=0$ and $\varepsilon^*_{\bm{K}\sigma}\cdot\varepsilon_{\bm{K}\sigma'}=-\delta_{\sigma\sigma'}$, for $\sigma,\sigma'=1,2$. Moreover, $j^{(++)}_{\bm{p}_{\mathrm{f}} \lambda_{\mathrm{f}},\bm{p}_{\mathrm{i}}\lambda_{\mathrm{i}}}(x)$ is the matrix element of the electron current operator with its $\nu$-component equal to
\begin{equation}
[j^{(++)}_{\bm{p}_{\mathrm{f}} \lambda_{\mathrm{f}},\bm{p}_{\mathrm{i}}\lambda_{\mathrm{i}}}(x)]^{\nu}
=\bar{\psi}^{(+)}_{\bm{p}_{\mathrm{f}} \lambda_{\mathrm{f}}}(x)\gamma^\nu \psi^{(+)}_{\bm{p}_{\mathrm{i}}\lambda_{\mathrm{i}}}(x).
\end{equation}
Here, $\psi^{(+)}_{\bm{p}\lambda}(x)$ is the Volkov solution of the Dirac equation coupled to the electromagnetic field \cite{Volkov}.

The Volkov solution of the Dirac equation for electrons of the four-momentum $p=(p^0,\bm{p})$, $p\cdot p=m^2_{\mathrm{e}}c^2$, is of the from
\begin{equation}
\psi^{(+)}_{\bm{p}\lambda}(x)=\sqrt{\frac{m_{\mathrm{e}}c^2}{VE_{\bm{p}}}}\Bigl[ 1-\frac{e}{2k\cdot p}\slashed{A}(k\cdot x)\slashed{k}\Bigr] u^{(+)}_{\bm{p}\lambda}\mathrm{e}^{-\mathrm{i}S^{(+)}_p(x)},
\label{volk1}
\end{equation}
where
\begin{equation}
S^{(+)}_p(x)=p\cdot x+\int_0^{k\cdot x}\Bigl[e\frac{A(\phi)\cdot p}{k\cdot p}-e^2\frac{A^2(\phi)}{2k\cdot p}\Bigr]\mathrm{d}\phi,
\label{volk2}
\end{equation}
$E_{\bm{p}}=cp^0$ and $u^{(+)}_{\bm{p}\lambda}$ is the free-electron bispinor normalized such that 
$\bar{u}^{(+)}_{\bm{p}\lambda}u^{(+)}_{\bm{p}\lambda'}=\delta_{\lambda\lambda'}$ with $\lambda=\pm$ labeling the spin degrees of freedom. 
The electromagnetic vector potential $A(k\cdot x)$ is assumed to be an arbitrary function of $k\cdot x$, with $k\cdot k=0$ and $\omega=ck^0$. 
Let the laser pulse lasts for time $T_{\mathrm{p}}$. Therefore, by choosing $\omega=2\pi/T_{\mathrm{p}}$ we can assume that $A(k\cdot x)$ 
vanishes for $k\cdot x<0$ and $k\cdot x>2\pi$. This allows to interpret the label $\bm{p}$ of the Volkov wave~\eqref{volk1} as the electron momentum in the remote past/future.

In analogy to the Bloch theorem in solid state physics, we can introduce the electron quasi-momentum $\bar{p}$. It describes the electron dressing 
by the electromagnetic field,
\begin{equation}
S_p(x)=\bar{p}\cdot x+ G_p(k\cdot x),
\label{volk3}
\end{equation}
where, for the most general, elliptically polarized plane-wave-fronted pulse,
\begin{equation}
A(k\cdot x)=A_0[\varepsilon_1f_1(k\cdot x)+\varepsilon_2f_2(k\cdot x)],
\label{volk5}
\end{equation}
the laser-dressed momentum has the form~\cite{KKcompton}
\begin{align}
\bar{p}=p-&\mu m_{\mathrm{e}}c\Bigl(\frac{p\cdot\varepsilon_1}{p\cdot n}\langle f_1\rangle+\frac{p\cdot\varepsilon_2}{p\cdot n}\langle f_2\rangle\Bigr)n \nonumber \\
+&\frac{1}{2}(\mu m_{\mathrm{e}}c)^2\frac{\langle f^2_1\rangle+\langle f^2_2\rangle}{p\cdot n}\, n.
\label{volk4}
\end{align}
Here, the parameter $\mu$ is defined according to Eq.~\eqref{kn5}.
The so-called shape functions, $f_j(k\cdot x)$, $j=1,2$, are arbitrary functions with continuous second derivatives that vanish outside the interval $(0,2\pi)$. 
For any of such functions $F(\phi)$, we define
\begin{equation}
\langle F\rangle=\frac{1}{2\pi}\int_0^{2\pi} F(\phi)\mathrm{d}\phi.
\label{volk6}
\end{equation}
Note that the pulses with plane wavefronts, which are considered in this paper, very well describe the interaction of laser fields with energetic electrons. 
This is provided that the kinetic energy of electrons is much larger than their ponderomotive energy in the laser field (see, e.g.~\cite{Lee2010}).

Our definition of the laser-dressed momentum~\eqref{volk4} follows directly from the Volkov solution~\eqref{volk1}. For the plane wave, 
the polarization-dependent terms in Eq.~\eqref{volk4} vanish, since $\langle f_j\rangle =0$. Note that the laser-dressed momentum, as the quantity defined in the laser field, 
cannot be a physical observable. It follows, however, from Eqs.~\eqref{volk1} and~\eqref{volk3} that the difference $(\bar{p}_{\mathrm{f}}-\bar{p}_{\mathrm{i}})$ (up to a four-vector proportional to $k$)
can be directly measured in an experiment, as it uniquely determines the Compton frequency $\omega_{\bm{K}}$ [see, Eq.~\eqref{comptonfreq} below]. This means that, in principle, we can redefine 
the dressed momentum \eqref{volk4} by adding an arbitrary four-vector, that is independent of $p$ and vanishes outside the laser pulse. By further assuming that this  
four-vector is not space- and time-dependent, and that it should be determined only by the four-vectors present in the definition of the laser pulse, one can consider the following modification \cite{KKbw},
\begin{equation}
\bar{p}\rightarrow\bar{p}+g_1\varepsilon_1+g_2\varepsilon_2+g_0 k,
\label{volk7}
\end{equation}
with $g_j=0$ in the absence of the laser field. It appears that this dressed momentum is on the mass-shell (i.e., $\bar{p}\cdot\bar{p}$ is independent of $p$) for a particular choice of $g_j$,
\begin{equation}
g_1=\mu m_{\mathrm{e}}c\langle f_1\rangle, \quad g_2=\mu m_{\mathrm{e}}c\langle f_2\rangle, \quad g_0=0,
\label{volk8}
\end{equation}
for which
\begin{equation}
\bar{p}\cdot\bar{p}=(\bar{m}_{\mathrm{e}}c)^2=(m_{\mathrm{e}}c)^2\Bigl[1+\frac{2U}{m_{\mathrm{e}}c^2}\Bigr],
\label{volk9}
\end{equation}
where $\bar{m}_{\mathrm{e}}$ is called the electron dressed mass, and
\begin{equation}
U=\frac{1}{2}\mu^2  m_{\mathrm{e}}c^2[\langle f^2_1\rangle-\langle f_1\rangle^2+\langle f^2_2\rangle-\langle f_2\rangle^2] .
\label{volk10}
\end{equation}
This result has led the authors of~\cite{HHIM} to the conclusion that the electron mass shift in a laser field could be measured by comparing 
the spectrum of Compton radiation induced by two different pulses, but of the same energy. However, it follows from our analysis that the electron 
mass shift can be well-defined \textit{only} for the particular choice of the momentum dressing [Eqs.~\eqref{volk7} and \eqref{volk8}].
This can create doubts about the physical nature of this quantity. 

Indeed, the quantity defined in Eq.~\eqref{volk10} is the direct generalization of the ponderomotive energy \eqref{kn4} in the electron reference frame for the finite laser pulses; in the arbitrary reference frame the ponderomotive energy can be defined as the time-component of the ponderomotive four-momentum $u$ multiplied by the speed of light $c$, i.e., as $U=cu^0$, where (cf., Ref.~\cite{KKphase} for the case when $\langle A\rangle=0$),
\begin{equation}
u^{\nu}=-\frac{e^2}{2}\,\frac{\langle A\cdot A\rangle-\langle A\rangle\cdot\langle A\rangle}{p\cdot k}\, k^{\nu} .
\label{volk11}
\end{equation}
This allows to define the dressed mass in the relativistically invariant form \cite{ReissPonder},
\begin{equation}
\bar{m}_{\mathrm{e}}^2=\frac{1}{c^2}(p+u)^2,
\label{volk12}
\end{equation}
which is independent of both the electron momentum $p$ and the fundamental frequency $\omega$, as well as of the laser pulse polarization vectors $\varepsilon_j$. 
Eq.~\eqref{volk12}, among others, has lead Reiss~\cite{ReissPonder} to the critique of the concept of the electron mass shift in a laser field. 
Note that such a dressing does not follow from the prescription~\eqref{volk7} if the parameters $g_j$ are $p$-independent and, in general, 
cannot be related to the `quasi-momentum' for the Volkov solution.

It is not the purpose of this paper to take part in the discussion concerning the mass shift. However, independently of the physical interpretation, 
both the mass shift and the ponderomotive momentum are uniquely defined by particular laser pulse characteristics; namely, by 
$\mu^2(\langle f^2_1\rangle+\langle f^2_2\rangle)$ and $\mu\langle f_j\rangle$ for $j=1,2$. It appears that also the Compton photon frequency 
depends on them (see, e.g., Eqs.~(29) and (30) in Ref.~\cite{KKscaling}). This indicates that there should exist an experimental method 
which allows to determine these parameters from direct measurements of the frequency 
spectrum of Compton photons. This would allow to determine either mass shift or the ponderomotive momentum of the electron in the laser field. It could be also 
used for the analysis of the peak intensity, the carrier-envelope phase or polarization properties of very intense, short laser pulses.

In this context, we further analyze the probability amplitude for the Compton process \eqref{ComptonAmplitude} and rewrite it as
\begin{equation}
{\cal A}(e^-_{\bm{p}_\mathrm{i}\lambda_\mathrm{i}}\longrightarrow e^-_{\bm{p}_\mathrm{f}\lambda_\mathrm{f}}+\gamma_{\bm{K}\sigma})
=\mathrm{i}\sqrt{\frac{2\pi\alpha c(m_{\mathrm{e}}c^2)^2}{E_{\bm{p}_\mathrm{f}}E_{\bm{p}_\mathrm{i}}\omega_{\bm{K}}V^3}}\, \mathcal{A}, \label{ct1}
\end{equation}
where 
\begin{align}
\mathcal{A}= \int\mathrm{d}^{4}{x}\, &  \bar{u}^{(+)}_{\bm{p}_\mathrm{f}\lambda_\mathrm{f}}\Bigl(1-\mu\frac{m_\mathrm{e}c}{2p_\mathrm{f}\cdot k}f(k\cdot x)\slashed{\varepsilon}\slashed{k}\Bigr)\slashed{\varepsilon}_{\bm{K}\sigma}  \nonumber \\
\times &\Bigl(1+\mu\frac{m_\mathrm{e}c}{2p_\mathrm{i}\cdot k} f(k\cdot x)\slashed{\varepsilon}\slashed{k}\Bigr) u^{(+)}_{\bm{p}_\mathrm{i}\lambda_\mathrm{i}}\,\mathrm{e}^{-\mathrm{i}S(x)},  \label{ct2}
\end{align}
with $S(x) = S^{(+)}_{{p}_\mathrm{i}}(x)-S^{(+)}_{{p}_\mathrm{f}}(x)-K\cdot x$. It has been shown in Ref.~\cite{KKcompton} that, 
by applying the Boca-Florescu transformation \cite{BocaFlorescu}, 
the frequency-angular distribution of energy of the emitted photons for an unpolarized and monoenergetic electron beam is given by
\begin{equation}\label{copton:spectrum:3}
\frac{{\rm d^{3}}E_{\rm C}}{{\rm d}\omega_{\bm K}{\rm d^{2}}\Omega_{\bm K}}
=\frac{1}{2}\sum_{\sigma=1,2}\sum_{\lambda_{\rm i}=\pm}\sum_{\lambda_{\rm f}=\pm}
\frac{{\rm d^{3}}E_{{\rm C},\sigma}(\lambda_{\rm i},\lambda_{\rm f})}{{\rm d}\omega_{\bm K}{\rm d^{2}}\Omega_{\bm K}}.
\end{equation}
Here,
\begin{equation}\label{copton:spectrum:2}
\frac{{\rm d^{3}}E_{{\rm C},\sigma}(\lambda_{\rm i},\lambda_{\rm f})}{{\rm d}\omega_{\bm K}{\rm d^{2}}\Omega_{\bm K}}
=\alpha\left|\mathcal{A}_{{\rm C},\sigma}(\omega_{\bm{K}},\lambda_{\rm i},\lambda_{\rm f})\right|^2,
\end{equation}
where the scattering amplitude equals
\begin{align}
\mathcal{A}_{{\rm C},\sigma}(\omega_{\bm{K}},\lambda_{\rm i},\lambda_{\rm f}) 
=&\frac{m_{\rm e}c K^{0}}{2\pi\sqrt{p_{\rm i}^{0}k^{0}(k\cdot p_{\rm f})}} \nonumber \\
\times &\sum_{N}D_{N}\frac{1-{\rm e}^{-2\pi\mathrm{i}(N-N_{\rm eff})}}{\mathrm{i}(N-N_{\rm eff})},
\label{copton:spectrum:1}
\end{align}
with $N_{\rm eff} = (K^{0}+\bar{p}^{0}_{\rm f}-\bar{p}^{0}_{\rm i})/k^{0}$. Moreover, the functions $D_N$ are defined in Ref.~\cite{KKcompton} by Eqs.~(23) and (44). 
In the equations above, the Compton frequency is related to the initial and final electron momenta through the conservation relations,
\begin{equation}
(\bar{p}_{\mathrm{i}}-\bar{p}_{\mathrm{f}}-K)\cdot n=0,\quad \bar{\bm{p}}^{\bot}_{\mathrm{i}}-\bar{\bm{p}}^{\bot}_{\mathrm{f}}-\bm{K}^{\bot}=\bm{0},
\label{conservation}
\end{equation}
where, for an arbitrary four-vector $a$, we define $\bm{a}^{\bot}=\bm{a}-(\bm{a}\cdot\bm{n})\bm{n}$. As a consequence, the Compton frequency,
\begin{eqnarray}
\omega_{\bm{K}}&=& c\frac{(K\cdot n)^2+(\bm{K}^{\bot})^2}{2K\cdot n} \nonumber \\
&= & c\frac{[(\bar{p}_{\mathrm{i}}-\bar{p}_{\mathrm{f}})\cdot n]^2+(\bar{\bm{p}}^{\bot}_{\mathrm{i}}-\bar{\bm{p}}^{\bot}_{\mathrm{f}})^2}{2(\bar{p}_{\mathrm{i}}-\bar{p}_{\mathrm{f}})\cdot n},
\label{comptonfreq}
\end{eqnarray}
is uniquely defined by the difference of dressed momenta, ($\bar{p}_{\mathrm{f}}-\bar{p}_{\mathrm{i}}$). 
For this reason one can select any form of the electron momentum dressing. Below, we adopt our definition, Eq.~\eqref{volk4}, 
as for such a choice the following equations hold: $\bar{p}\cdot n=p\cdot n$, $\bar{p}\cdot\varepsilon_j=p\cdot\varepsilon_j$ 
and $\bar{\bm{p}}^{\bot}=\bm{p}^{\bot}$, that significantly simplify analytical calculations.

In the following, we shall consider the linearly polarized laser pulse such that, for $0\leqslant \phi=k\cdot x\leqslant 2\pi$, the vector potential has 
a general form $A(\phi)=A_0 \varepsilon f(\phi)$ [i.e., we put $\varepsilon=\varepsilon_1$, $f(\phi)=f_1(\phi)$ and $f_2(\phi)=0$] and the electric field vector 
equals $\bm{\mathcal{E}}(\phi)=-\omega A_0 \bm{\varepsilon} f^{\prime}(\phi)$. The shape function $f(\phi)$ is defined via its derivative,
\begin{equation}
f'(\phi)=\begin{cases} 0, & \phi <0, \cr
                 N^{\prime}_f\sin^2\bigl(N_{\mathrm{rep}}\frac{\phi}{2}\bigr)\sin(N_{\mathrm{rep}}N_{\mathrm{osc}}\phi), & 0\leqslant\phi\leqslant 2\pi,\cr
								0, & \phi > 2\pi,
			  \end{cases}
\label{t2}
\end{equation}
where we assume that $f(0)=0$. Above, the integers $N_{\mathrm{rep}}$ and $N_{\mathrm{osc}}$ determine the number of identical pulses in a train 
and the number of cycles in each pulse, respectively, whereas $N^{\prime}_f$ is a suitably chosen normalization constant. Since the duration 
of the laser pulse is $T_{\mathrm{p}}$, we can also define the fundamental frequency, $\omega=2\pi/T_{\mathrm{p}}$, and the central one, 
$\omega_{\mathrm{L}}=N_{\mathrm{rep}}N_{\mathrm{osc}}\omega$ of the laser field, which is supposed to be fixed and equal to 
$\omega_{\mathrm{L}}=1.55\mathrm{eV}\approx 3\times 10^{-6}m_{\mathrm{e}}c^2$ in all calculations presented here. Moreover, 
in the following we assume that $N^{\prime}_f=N_{\mathrm{rep}}N_{\mathrm{osc}}$, which guarantees that the time-averaged intensity 
of the laser pulse is independent of $N_{\mathrm{rep}}$ and $N_{\mathrm{osc}}$, as for this particular selection the amplitude 
of the electric field scales as $\omega_{\mathrm{L}}\mu$. Note that, for the rectangular pulse, the $\sin^2$ envelope is not 
present in~\eqref{t2}, meaning that the laser pulse depends only on the product $N_{\mathrm{rep}}N_{\mathrm{osc}}$.

\begin{figure}
\includegraphics[width=6.5cm]{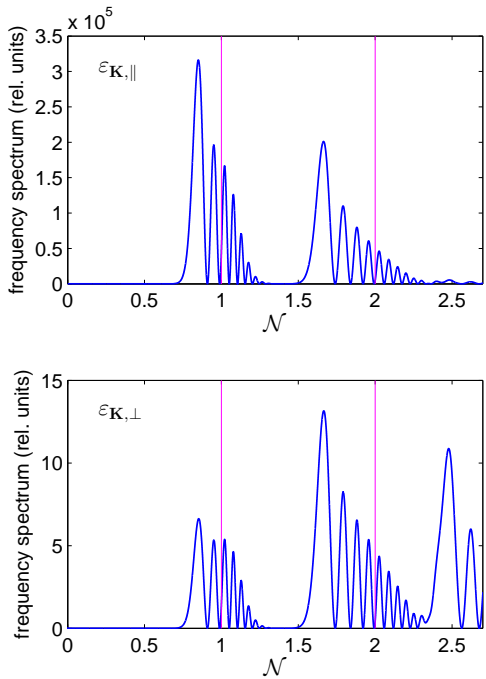}%
\caption{(Color online) Shows the spectra of Compton radiation [Eq.~\eqref{copton:spectrum:3}] resulting from the head-on collision of the linearly 
polarized laser pulse and an electron of momentum $\bm{p}_{\mathrm{i}}=-10^3m_{\mathrm{e}}c\bm{e}_z$. The laser pulse ($\mu=1$ 
and $\omega_{\mathrm{L}}=1.55\mathrm{eV}$) propagates along the $z$ direction and is polarized along the $x$ direction. 
These two axes determine the scattering plane. The pulse has a $\sin^2$ envelope [Eq.~\eqref{t2}] with $N_{\mathrm{osc}}=30$ and 
$N_{\mathrm{rep}}=1$. The Compton photon is emitted in the direction of $\theta_{\bm{K}}=0.9999\pi$ and $\varphi_{\bm{K}}=\pi$, 
with the polarization vector either parallel (upper panel) or perpendicular to the scattering plane (lower panel). 
The energy spectra are presented as functions of $\mathcal{N}$, where the vertical lines mark the integer values of this argument.
\label{long1case21c20150228}}
\end{figure}

In analogy to the original Klein-Nishina formula [Eqs.~\eqref{kn2} and~\eqref{kn9}], we present the frequency spectrum as a function of $\mathcal{N}$,
\begin{equation}
\mathcal{N}=\frac{\omega_{\bm{K}}\omega_{\mathrm{cut}}}{\omega_{\mathrm{L}} (\omega_{\mathrm{cut}}-\omega_{\bm{K}})}
\Bigl(\frac{p_{\mathrm{i}}\cdot n_{\bm{K}}}{p_{\mathrm{i}}\cdot n}+\frac{U}{c}\frac{n\cdot n_{\bm{K}}}{p_{\mathrm{i}}\cdot n}\Bigr).
\label{kn10}
\end{equation}
In this case, we expect that in the limit of a very long pulse the peaks in the spectrum will appear for $\mathcal{N}$'s very close to integers, as it indeed takes place for a monochromatic plane wave.

In Fig.~\ref{long1case21c20150228}, we present the angular-resolved distributions of Compton radiation [Eq.~\eqref{copton:spectrum:3}] 
generated by a single pulse, with $N_{\mathrm{osc}}=30$ field oscillations. This distribution is presented as the function of $\mathcal{N}$. 
One could expect that, for such a long driving pulse, the Klein-Nishina formula could be successfully used and, that the dominant peaks would
correspond to integer values of $\mathcal{N}$. Clearly, we do not observe such a behavior as the peaks are red-shifted, independently 
of the Compton photon polarization. As we will show in the next section, the situation is changed when a train of incident pulses is considered.

\section{Diffraction formula}
\label{sec:diffraction}

It was shown in~\cite{KKcomb2} that, for a finite train of pulses, the Compton probability amplitude has the diffraction-type form,
\begin{align}
\mathcal{A}_{\mathrm{C},\sigma}&(\omega_{\bm{K}},\lambda_{\mathrm{i}},\lambda_{\mathrm{f}})=\exp\Bigl[\mathrm{i}\Phi_{\mathrm{C},\sigma}(\omega_{\bm{K}},\lambda_{\mathrm{i}},\lambda_{\mathrm{f}})\Bigr] \nonumber \\
 &\times\frac{\sin(\pi\bar{Q}^+/k^0)}{\sin(\pi\bar{Q}^+/k^0N_{\mathrm{rep}})}|\mathcal{A}^{(1)}_{\mathrm{C},\sigma}(\omega_{\bm{K}},\lambda_{\mathrm{i}},\lambda_{\mathrm{f}})|,
\label{ci17}
\end{align}
where $\mathcal{A}^{(1)}_{{\rm C},\sigma}$ is the Compton amplitude for a single pulse and $\Phi_{\mathrm{C},\sigma}(\omega_{\bm{K}},\lambda_{\mathrm{i}},\lambda_{\mathrm{f}})$ 
is the Compton global phase. In the above equation $\bar{Q}^+=\bar{p}_{\mathrm{i}}^+-\bar{p}_{\mathrm{f}}^+-K^+$ where, for an arbitrary four-vector $a$, we define 
$a^+=a^0-(a\cdot n)/2= (a^0+\bm{a}\cdot\bm{n})/2$. For particular frequencies of emitted photons ($\omega_{\bm{K},N}$ with integer $N$), that satisfy the condition
\begin{equation}
\pi\bar{Q}^+=-\pi N N_{\mathrm{rep}}k^0,
\label{ci18}
\end{equation}
we observe the coherent enhancement of the Compton amplitude. This, in turn, leads to the quadratic, $N_{\mathrm{rep}}^2$, enhancement of the respective probability distribution. 
In contrast to the classical Thomson process, these frequencies are not \textit{exactly} equally spaced in the allowed frequency region, 
$0<\omega_{\bm{K}}<\omega_{\mathrm{cut}}$. When $\omega_{\bm{K}}$ approaches the cut-off value $\omega_{\mathrm{cut}}$ [Eq. \eqref{kn8}], i.e., when the quantum recoil of the scattered 
electron cannot be neglected, the spectrum of $\omega_{\bm{K},N}$ becomes increasingly denser. This means that one can generate the Compton-based 
frequency combs with equidistant peak frequencies only within limited frequency intervals.

The Compton global phase equals
\begin{equation}
\Phi_{\mathrm{C},\sigma}(\omega_{\bm{K}},\lambda_{\mathrm{i}},\lambda_{\mathrm{f}})=
-\pi\frac{\bar{Q}^+}{k^0}+\Phi_{\mathrm{C},\sigma}^{\mathrm{dyn}}(\omega_{\bm{K}},\lambda_{\mathrm{i}},\lambda_{\mathrm{f}}),
\label{ci20}
\end{equation}
where $\Phi_{\mathrm{C},\sigma}^{\mathrm{dyn}}$ is the so-called dynamic phase~\cite{KKcomb2}. For arbitrary laser pulses and polarizations 
of emitted photons, the dynamic phase can only be calculated numerically. It happens that, for pulses considered in this paper, the dynamic phase is independent 
of $\omega_{\bm{K}}$. This means that, for frequencies $\omega_{\bm{K},N}$ satisfying the condition~\eqref{ci18}, the global phase is equal to 
\begin{equation}
\Phi_{\mathrm{C},\sigma}(\omega_{\bm{K},N},\lambda_{\mathrm{i}},\lambda_{\mathrm{f}})=
\pi N\,N_{\mathrm{rep}}+\Phi_{\mathrm{C},\sigma}^{\mathrm{dyn}}(\omega_{\bm{K},N},\lambda_{\mathrm{i}},\lambda_{\mathrm{f}}),
\label{ci21}
\end{equation}
and, hence, it takes on the same values modulo $\pi$. The selection of these particular phases for the peak frequencies 
leads not only to the enhancement of the frequency spectrum, but also to the synthesis of ultra-short pulses of radiation generated during the Compton scattering~\cite{KKsuper}.

\begin{figure}
\includegraphics[width=6.5cm]{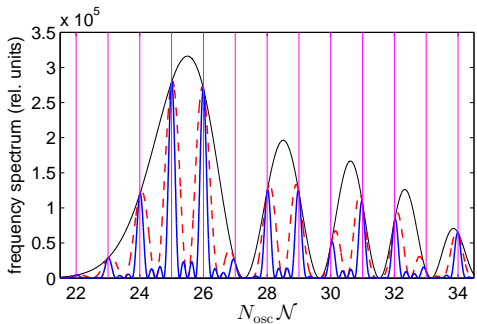}%
\caption{(Color online) The same as in Fig.~\ref{long1case21c20150228} but only for the Compton photon polarized in the scattering plane. 
We show the spectra for $N_{\mathrm{rep}}=1$ (black envelope), $N_{\mathrm{rep}}=2$ (dashed red line), and $N_{\mathrm{rep}}=4$ (solid blue line). 
The vertical lines mark the integer values of $N_{\mathrm{osc}}\mathcal{N}$, for which we observe the approximate positions of peaks for $N_{\mathrm{rep}}>1$.
\label{long124case21c20150228}}
\end{figure}

As an illustration, we consider the same laser pulse as in Fig.~\ref{long1case21c20150228} but repeated $N_{\mathrm{rep}}$ times. In Fig.~\ref{long124case21c20150228}, 
we show the Compton spectra for $N_{\mathrm{rep}}=1$, 2, and 4, when divided by $N_{\mathrm{rep}}^2$. The results are presented as the functions of $N_{\mathrm{osc}}\mathcal{N}$. 
The clearly visible peaks occur for $N_{\mathrm{rep}}>1$. Their positions are almost independent of $N_{\mathrm{rep}}$ and they correspond to the integer values of 
$N_{\mathrm{osc}}\mathcal{N}$. Note that $\omega=\omega_{\mathrm{L}}/N_{\mathrm{osc}}$ is the fundamental frequency of the individual laser pulse from the train. 
Such a pulse can be approximately interpreted as a coherent superposition of at most $N_{\mathrm{osc}}$ photon states of frequencies $K\omega$, $K=1,\ldots,N_{\mathrm{osc}}$ 
(of course, photons with other frequencies are also present, but with smaller amplitudes). Therefore, in the course of the Compton scattering, the electron can absorb these 
photons with the total energy $N\omega$ and emit a single photon of frequency $\omega_{\bm{K},N}$. However, due to the time-frequency uncertainty relation 
and an incoherent interference of probability amplitudes, the spectrum is smeared out such that it is not possible to clearly prescribe peaks to orders $N$.
This is clearly seen in Figs.~\ref{long1case21c20150228} and~\ref{long124case21c20150228}. The situation changes if we consider the sequence of at least two such pulses. 
Now, due to the constructive interference, the processes with integer $N_{\mathrm{osc}}\mathcal{N}$ are coherently enhanced. As a result, we observe in the spectrum 
the clearly resolved peaks already for $N_{\mathrm{rep}}=2$.

The coherent enhancement of the Compton frequency spectra does not take place for a single pulse with the time-varying envelope. It appears, however, that 
important features of a single pulse can be precisely determined from positions of the main diffraction peaks in the Compton spectrum,
when generated by a train of such pulses.

\section{Generalized Klein-Nishina formula}
\label{sec:gklein}

As we have demonstrated above, the application of the pulse train with two subpulses already allows to increase the resolution of the frequency spectrum of Compton radiation 
such that one can unambiguously prescribe an integer number to the individual peaks. We have shown this for a long pulse ($N_{\rm osc}=30$), for which $\langle f\rangle$ is negligibly small, 
so that the original Klein-Nishina formula may be applied. For shorter pulses, $\langle f\rangle$ starts to be significantly different than zero and the generalization 
of the Klein-Nishina formula, that accounts for this fact, is necessary. We apply the diffraction formula and determine the Compton photon frequency 
by solving the system of equations~\eqref{conservation} and~\eqref{ci18}. After some algebra, we arrive at the following GKN formula valid 
for a pulse train of an arbitrary polarization,
\begin{equation}
\omega_{\bm{K},N}=\frac{(N/N_{\mathrm{osc}})\omega_{\mathrm{L}}}{\frac{p_{\mathrm{i}}\cdot n_{\bm{K}}}{p_{\mathrm{i}}\cdot n}+\frac{\nu n\cdot n_{\bm{K}}+g_1p_{\mathrm{i},1}+g_2p_{\mathrm{i},2}}{(p_{\mathrm{i}}\cdot n)^2}+\frac{(N/N_{\mathrm{osc}})\omega_{\mathrm{L}}}{\omega_{\mathrm{cut}}}}.
\label{gkn1}
\end{equation}
Here, $g_1$ and $g_2$ are defined in Eq.~\eqref{volk8},
\begin{equation}
\nu=\frac{1}{2}(\mu m_{\mathrm{e}}c)^2(\langle f_1^2\rangle + \langle f_2^2\rangle),
\label{gkn2}
\end{equation}
and (for $j=1,2$)
\begin{equation}
p_{\mathrm{i},j}=(p_{\mathrm{i}}\cdot n)(n_{\bm{K}}\cdot \varepsilon_j)-(p_{\mathrm{i}}\cdot \varepsilon_j)(n\cdot n_{\bm{K}}).
\label{gkn3}
\end{equation}
As for the original Klein-Nishina formula, the quantum signature is hidden in the definition of $\omega_{\mathrm{cut}}$ [Eq.~\eqref{kn8}]. 
The frequencies determined by Eq.~\eqref{gkn1} mark the positions of main peaks in the Compton spectrum. Similarly, 
one can find frequencies of the secondary peaks (if $N_{\mathrm{rep}}>2$) and zeros (if $N_{\mathrm{rep}}>1$) 
in the angular-resolved frequency distributions. It is worth noting that now the polarization-dependent terms 
appear not in the unphysical dressing of the electron initial and final momenta, but in the definition of the directly measurable 
quantity; in other words, they affect the peak frequencies of the Compton spectrum. Similarly to the original Klein-Nishina formula we define the quantity
\begin{align}
\mathcal{N}_{\mathrm{GKN}}=&\frac{N_{\mathrm{osc}}\omega_{\bm{K}}\omega_{\mathrm{cut}}}{\omega_{\mathrm{L}} (\omega_{\mathrm{cut}}-\omega_{\bm{K}})}
\Bigl(\frac{p_{\mathrm{i}}\cdot n_{\bm{K}}}{p_{\mathrm{i}}\cdot n} \nonumber \\
+&\frac{\nu n\cdot n_{\bm{K}}+g_1p_{\mathrm{i},1}+g_2p_{\mathrm{i},2}}{(p_{\mathrm{i}}\cdot n)^2}\Bigr),
\label{gkn4}
\end{align}
which acquires integer values for peak frequencies $\omega_{\bm{K},N}$. For $\langle f_j\rangle=0$ and $N_{\mathrm{osc}}=1$, this formula reduces 
to the original one, given by Eq.~\eqref{kn10}.

\begin{figure}
\includegraphics[width=6.5cm]{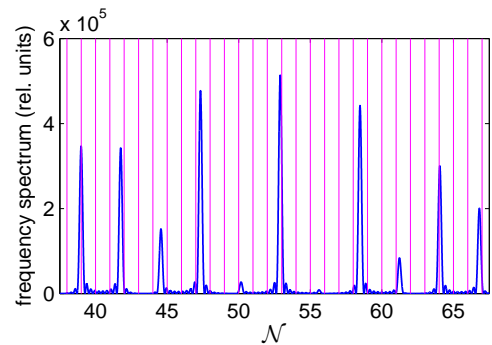}%
\caption{(Color online) The same as in Fig.~\ref{long1case21c20150228} but for the rectangular pulses with $N_{\mathrm{osc}}=1$ and $N_{\mathrm{rep}}=10$. 
The laser field intensity is characterized by the parameter $\mu=3$. The Compton photon is emitted in the direction specified by $\theta_{\bm{K}}=0.999\pi$
and $\varphi_{\bm{K}}=\pi$. The energy spectrum is presented as the function of $\mathcal{N}$ [Eq.~\eqref{kn10}].
\label{plane10case21s20150228}}
\end{figure}

The derivation of the formula~\eqref{ci17} shows that, in order to observe a coherent enhancement of the Compton spectra, the modulations of the driving pulse cannot be arbitrary~\cite{KKcomb2}. 
Both the integral of the electric field over the time duration of a single modulation and the vector potential in the beginning and at 
the end of it have to vanish. In other words, we have to deal with a train of pulses. This precludes the application of the diffraction 
formula not only to single pulses with varying in time envelopes, but also to the rectangular pulses of an arbitrary polarization. 
The exception is the linearly polarized rectangular pulse, but even in this case one has to redefine the vector potential such that 
$\langle f\rangle\neq 0$. This means that, irrespectively of the duration of such a rectangular pulse, the original Klein-Nishina formula~\eqref{kn2} 
is not applicable, unless the particular geometry is selected such that $p_{\mathrm{i},j}=0$ for $g_j\neq 0$. For instance, in the electron 
reference frame this happens if $\bm{n}_{\bm{K}}\cdot\bm{\varepsilon}_j=0$, as $p_{\mathrm{i}}\cdot\varepsilon_j=0$. This corresponds to the case
when the Compton photon is ejected perpendicular to the laser field polarization vector.

\begin{figure}
\includegraphics[width=6.5cm]{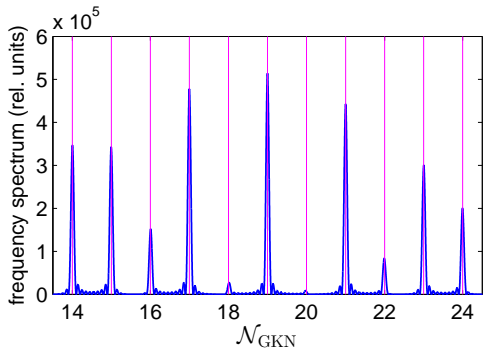}%
\caption{(Color online) The same as in Fig.~\ref{plane10case21s20150228}, but the spectrum is plotted as a function of $\mathcal{N}_{\mathrm{GKN}}$ [Eq.~\eqref{gkn4}].
\label{plane10gcase21s20150228}}
\end{figure}

In Figs.~\ref{plane10case21s20150228} and~\ref{plane10gcase21s20150228}, we consider a generic case when $p_{\mathrm{i},j}\neq 0$ for the linearly polarized laser field. 
In these figures we compare the same spectrum of emitted radiation, but we present it as the function of either $\mathcal{N}$ or $\mathcal{N}_{\mathrm{GKN}}$.
We see that the main peaks of this spectrum correspond exactly to the integer values of $\mathcal{N}_{\mathrm{GKN}}$. We also observe a dramatic 
difference in numerical values for $\mathcal{N}$ and $\mathcal{N}_{\mathrm{GKN}}$. For instance, the first peak corresponds to $\mathcal{N}=39$ 
or $\mathcal{N}_{\mathrm{GKN}}=14$. If the original Klein-Nishina formula had been used for interpreting this peak one would ascribe it to 
the process with absorption of 39 laser photons (not accounting for the fact that the majority of the peaks could not be interpreted this 
way, since they do not match the integer values of $\mathcal{N}$). In contrast, for the GKN formula, all the main peaks 
can be interpreted as the result of absorption of an integer number of laser photons. Note that the positions of these peaks are the same with 
the increasing  $N_{\mathrm{rep}}$, whereas their widths become increasingly narrower.

\section{Few-cycle laser pulses}
\label{sec:short}

A closer look at Fig.~\ref{long124case21c20150228} shows that for $N_{\mathrm{rep}}>1$ some peaks in the spectrum do not scale as $N_{\mathrm{rep}}^2$. 
This concerns peaks located close to the frequencies for which the distribution for the single pulse vanishes. As one can see, after dividing these 
distributions by $N_{\mathrm{rep}}^2$, the spectrum for $N_{\mathrm{rep}}=1$ represents the envelope for the main diffraction peaks (i.e., observed for $N_{\mathrm{rep}}>1$).
This means that the spectra are tangent to each other for frequencies close to the main peaks. If the Compton spectrum for a single pulse shows rapid modulations and 
the diffraction peak is located at the edge of a particular modulation, then the peak frequency does not correspond to the one for which the spectra are tangent. 
In these cases, it may happen that the application of a pulse train does not enhance, but rather suppresses, the generated radiation. This is the reason why 
in Fig.~\ref{plane10gcase21s20150228} some of the diffraction peaks are hardly visible. To avoid the suppression of emitted radiation, it 
is advisable to use such laser pulses, or such scattering kinematics, that the spectrum originating from a single pulse exhibits a broad structure; the so-called supercontinuum.
Note that the formation of supercontinua was discussed recently in the context of Thomson and Compton scattering, and the synthesis of zepto- and yoctosecond pulses of radiation~\cite{KKsuper}. 
It appears that the best choice for their generation is to use few-cycle laser pulses. We shall illustrate this below for $N_{\mathrm{osc}}=1$ and 3.

\begin{figure}
\includegraphics[width=6.5cm]{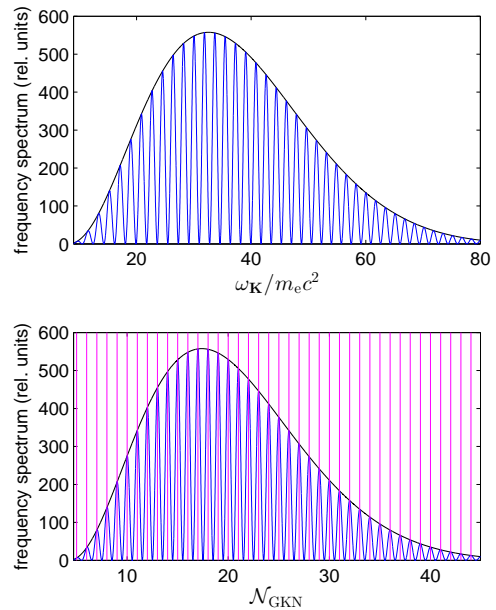}%
\caption{(Color online) The same as in Fig.~\ref{long1case21c20150228}, but for $N_{\rm osc}=1$ and only for the Compton photon polarized in the scattering plane. 
The curves in each panel represent two cases: $N_{\mathrm{rep}}=1$ (smooth envelope) and $N_{\mathrm{rep}}=2$ (densely distributed peaks). In the upper panel 
the energy spectra are presented as functions of the Compton photon frequency $\omega_{\bm{K}}$, whereas in the lower frame as functions of 
$\mathcal{N}_{\mathrm{GKN}}$. The vertical lines in the lower panel mark the integer values of ${\cal N}_{\rm GKN}$, that exactly coincide with the positions of peaks. 
The distance between the peaks is roughly $1.79m_{\mathrm{e}}c^2$.
\label{single12gcase21s20150228}}
\end{figure}
\begin{figure}
\includegraphics[width=6.5cm]{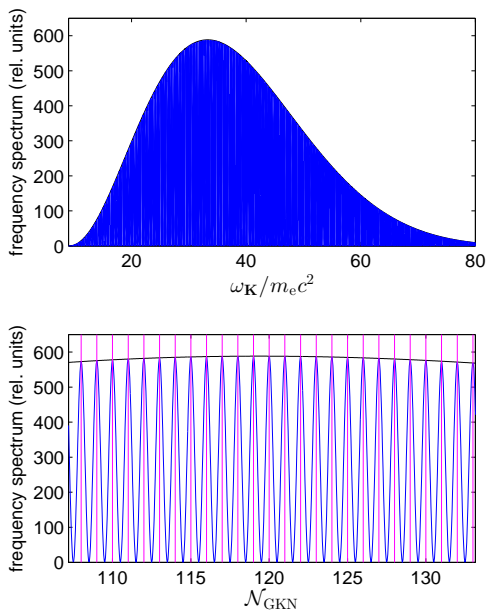}%
\caption{(Color online) The same as in Fig.~\ref{single12gcase21s20150228}, but for $N_{\mathrm{osc}}=3$. 
Since the peaks are denser, in the lower panel only part of the spectrum is presented. The peaks are separated by roughly $0.33m_{\mathrm{e}}c^2$.
\label{triple12gcase21s20150228}}
\end{figure}

In Fig.~\ref{single12gcase21s20150228}, we present the Compton spectrum induced by a single-cycle pulse. The spectrum consists of the broad 
supercontinuum which extends from nearly $10m_{\mathrm{e}}c^2$ up to $80m_{\mathrm{e}}c^2$. If we repeat this pulse, $N_{\mathrm{rep}}=2$, 
we observe the formation of the diffraction peaks for integer values of $\mathcal{N}_{\mathrm{GKN}}$ (lower panel), which proves the validity 
of the GKN formula. For larger values of $N_{\mathrm{rep}}$, the positions of the main peaks stay the same but their widths 
become more narrow. The energy separation between the adjacent peaks is nearly the same ($\sim 1.79m_{\mathrm{e}}c^2$). 
Note that the pulse train under consideration is the superposition of two plane waves of frequencies $\omega_{\mathrm{L}}$ and $2\omega_{\mathrm{L}}$. 
Also, the peak intensity of the laser field is not very large, as it does not exceed $10^{19}\mathrm{W/cm}^2$. This suggests that our theoretical 
predictions could be verified experimentally, for instance at the ELI facility~\cite{eli}.

For pulses with more oscillations, the situation is similar. In Fig.~\ref{triple12gcase21s20150228}, we show this for $N_{\mathrm{osc}}=3$. 
The only difference is that now the distribution of the diffraction peaks is denser, with the energy separation of roughly $0.33m_{\mathrm{e}}c^2$. 
As above, the main peaks (for $N_{\mathrm{rep}}=2$ there are only the main diffraction peaks and the weaker secondary ones show up for $N_{\mathrm{rep}}>2$, 
as presented in Fig.~\ref{long124case21c20150228}) correspond to the clearly prescribed integer values of $\mathcal{N}_{\mathrm{GKN}}$. This, again, 
proves the validity of the GKN formula derived in this paper.

It follows from Eq.~\eqref{gkn4} that, knowing the geometry of the Compton scattering and the electron initial energy, measuring frequencies of only 
three consecutive peaks in the spectrum (for $N_{\mathrm{rep}}>1$) leads to the determination of the two important parameters of linearly polarized 
pulses which comprise the train: $\mu^2\langle f^2\rangle$ and $\mu\langle f\rangle$. If the form of the envelope is known such measurements 
allow to determine the peak intensity of incident pulses, which is characterized by the parameter $\mu$. Another possibility is to map the positions of
the Compton peaks to the carrier envelope phase, assuming that the envelope type and the peak intensity is known.
Similar measurements for two different geometries can extract the values of $\mu^2(\langle f_1^2\rangle+\langle f_2^2\rangle)$ and $\mu\langle f_j\rangle$ ($j=1,2$) 
for elliptically polarized driving pulses and, hence, also their polarization properties.

\section{Conclusions}
\label{sec:conclusions}

We have demonstrated that, by using a train consisting of a finite number of identical pulses, one can generate the Compton radiation with well-resolved peaks.
In other words, we propose the mechanism to reduce the spectral broadening of the emitted radiation which typically occurs if a few-cycle pulse interacts with the electron
(see, for instance, Refs.~\cite{SKscale1,SKscale2,KKcompton,KKscaling,BocaFlorescu,Mackenroth}). 
This is a complementary proposal to the one presented in Ref.~\cite{Brad,Terzic,Seipt2015}, where a single but chirped initial pulse
was used in order to compensate for the spectral broadening. Based on this result, we have derived the generalized Klein-Nishina formula. 

The GKN formula~\eqref{gkn4} predicts the positions of well-resolved peaks in the Compton spectrum, when driven by a finite train of pulses.
We argue that, by analyzing the positions of the peaks in the frequency domain of Compton photons, it is possible to determine laser pulse parameters, 
$\mu^2(\langle f_1^2\rangle+\langle f_2^2\rangle)$ and $\mu\langle f_j\rangle$ for both linear polarizations. This means that the proposed method 
can be applied, for instance, to determine polarization properties of such pulses and either their peak intensity or their carrier-envelope phase. 
Note that a similar analysis can be carried out for other fundamental processes of strong-field quantum electrodynamics, like the laser-induced 
Breit-Wheeler and Bethe-Heitler pair creation. These possibilities are under investigations now.

\section*{Acknowledgements}

This work is supported by the Polish National Science Center (NCN) under Grant No. 2012/05/B/ST2/02547. 
Moreover, K.K. acknowledges the support from the Kosciuszko Foundation and F.C.V. acknowledges the Foundation 
for Polish Science International PhD Projects Programme co-financed by the EU European Regional Development Fund.

\end{document}